\documentclass[conference]{IEEEtran}

\makeatletter

\def\ps@IEEEtitlepagestyle{%
  \def\@oddfoot{\mycopyrightnotice}%
  \def\@evenfoot{}%
}
\def\mycopyrightnotice{%
  {\footnotesize U.S. Government work not protected by U.S. copyright.\hfill}
  \gdef\mycopyrightnotice{}
}

\usepackage{blindtext}
\usepackage{eso-pic}
\IEEEoverridecommandlockouts
\usepackage{cite}
\usepackage{amsmath,amssymb,amsfonts}
\usepackage{algorithmic}
\usepackage{graphicx}
\usepackage{textcomp}
\usepackage{xcolor}
\usepackage{tabularx}
\usepackage{booktabs}
\usepackage{listings} 
\def\BibTeX{{\rm B\kern-.05em{\sc i\kern-.025em b}\kern-.08em
    T\kern-.1667em\lower.7ex\hbox{E}\kern-.125emX}}
    
\usepackage{eso-pic}
\newcommand\AtPageUpperMyright[1]{\AtPageUpperLeft{%
 \put(\LenToUnit{0.17\paperwidth},\LenToUnit{-2cm}){%
     \parbox{0.9\textwidth}{\raggedleft\fontsize{8}{11}\selectfont #1}}%
 }}%
\newcommand{\conf}[1]{%
\AddToShipoutPictureBG*{%
\AtPageUpperMyright{#1}
}
}

\begin{document}
\title{\vspace*{1cm} Enhancing Code Quality with Generative AI: Boosting Developer Warning Compliance
}

\author{\IEEEauthorblockN{1\textsuperscript{st} Hansen Chang}
\IEEEauthorblockA{\textit{Electrical and Computer Engineering Department} \\
\textit{United States Naval Academy}\\
Annapolis, United States \\
m250924@usna.edu}
\and
\IEEEauthorblockN{2\textsuperscript{nd} Christian DeLozier}
\IEEEauthorblockA{\textit{Electrical and Computer Engineering Department} \\
\textit{United States Naval Academy}\\
Annapolis, United States \\
delozier@usna.edu}
}

\maketitle
\conf{\textit{ Paper was accepted to ICECET 2025 but could not be published due to credit card payment issues)}}

\begin{abstract}
Programmers have long ignored warnings, especially those generated by static analysis tools, due to the potential for false-positives.  In some cases, warnings may be indicative of larger issues, but programmers may not understand how a seemingly unimportant warning can grow into a vulnerability.  Because these messages tend to be long and confusing, programmers tend to ignore them if they do not cause readily identifiable issues. Large language models can simplify these warnings, explain the gravity of important warnings, and suggest potential fixes to increase developer compliance with fixing warnings.
\end{abstract}


\emph{The views expressed in this article are  those of the author(s) and do not reflect the official policy or position of the U.S. Naval Academy, Department of the Navy, the Department of Defense, or the U.S. Government.}

Warning messages generated by compilers and static analysis tools \cite{find-bugs} have historically been overlooked and ignored \cite{why-no-static}.  Compiler messages are considered to be difficult to understand due to complexity and scale \cite{errors-survey}, and static analysis warnings are often ignored due to the prevalence of false positives \cite{static-lessons}.

Warnings in static analysis can range from minor style issues to critical errors that could lead to system vulnerabilities. While they provide useful insights, research indicates that developers often ignore or suppress these warnings due to time constraints, perceived irrelevance, or warning fatigue. This tendency to overlook warnings can lead to technical debt and system instability over time. Improving the interpretability and clarity of warnings has been identified as a potential way to enhance developer compliance, which in turn strengthens software quality \cite{program-analysis-empirical}.

Large language models \cite{vaswani} can assist developers with understanding hard to read compiler messages.  Prior work has shown that large language models can effectively sift through large amounts of static warnings and prioritize important warnings for developers \cite{static-llm}. This study determined that analyzing static analysis messages with large language models was a valid strategy, with a 69-81\% accuracy depending on the scheme.  Another recent study done on enhancing compiler error messages concluded that these edited messages reduced the number of errors in students \cite{enhancing-compiler-warnings}. In this study, an editor called Decaf was designed and implemented to take in compiler errors and simply them. The simplified message was then given to student programmers in order to analyze the effectiveness.  Additional work has shown that large language models are capable of performing other useful tasks related to static analysis, such as translating static analysis tests developed for one programming language to another \cite{ignatyev-llm-static}.

\begin{figure}
    \centering
    \includegraphics[width=.48\textwidth]{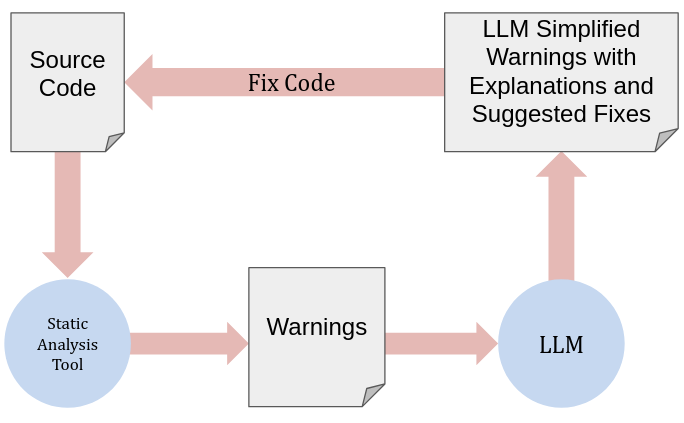}
    \caption{Overview of Using LLMs to Boost Developer Warning Compliance}
    \label{fig:overview}
\end{figure}

We intend to build on this prior work to utilize the growing capabilities of large language models in order to simplify static warnings and convince users to fix potentially dangerous bugs. Figure \ref{fig:overview} provides an overview of this approach in which warnings produced by the static analysis tool are fed into the LLM to produce a simplified message that includes rationale for fixing the bug and a suggested fix.  Our hypothesis is that given a plain language explanation of why a static analysis warning is important and a suggested fix that will prevent that warning, programmers will be more likely to fix code with potential issues.

\section{Experimental Setup}

In the following section, we describe our ongoing experiments with large language models in the context of encouraging developers to fix warnings in their code.  We describe the current and future benchmark suites that we intend to use, the static analysis tools that we have examined, our current efforts in prompt engineering, and our experience with a set of large language models in this context.

\subsection{Benchmark Suite}

The NIST Juliet Suite \cite{juliet} contains a variety of test cases, including cases that demonstrate major vulnerabilities and others that result in minor warnings.  For the scope of our research, we are attempting to measure the effectiveness of large language models in convincing programmers to correct minor warnings that might usually be ignored given that fixing these minor warnings often leads to directly or indirectly fixing the major vulnerability.

The code provided by the Suite is designed to run via commands in scripts and a common build environment in order to execute every single case to test the capabilities of bug finding tools. To re-purpose it for our needs, we take individual code files and trim the code down to the basic "good" and "bad" cases and add the necessary header files and edits needed for it to compile and run. As we continue to work on this problem, we intend to investigate other available static analysis benchmark suites like the Toyota ITC benchmarks \cite{itc} and larger static analysis examples from real applications.

Each test case was moved into a new C file to remove the need for the Juliet suite's supporting include files.  Variable and function names were obfuscated to remove textual hints such as "good" and "bad" from the code that would otherwise provide unrealistic assistance to the large language model in understanding the test cases.  Likewise, comments that textually indicated that a bug was present in the code were removed.  We intend to use these stripped-down test cases in a future user study of undergraduate programmers to determine how motivated to fix warnings and effective at fixing warnings they are given the assistance of a large language model.

\subsection{Static Analysis Tools}

For this work, we have examined the use of two popular static analysis tools for C and C++ code - cppcheck \cite{cppcheck} and clang-check \cite{clang-check}.  We ran cppcheck with the \texttt{--enable=all} and \texttt{inconclusive} flags to yield all possible static analysis warnings.  However, we suppressed warnings that are caused by the Juliet suite's execution framework.  For example, \texttt{missingOverride} warnings occur due to the Juliet test suite files not having an \texttt{override} specifier on virtual methods, and these bugs are not related to the errors demonstrated by the test cases.  We ran clang-check with the \texttt{--analyze} and \texttt{--fixit} flags.

\subsection{Prompt Engineering}

Prompt Engineering is constructing the right prompt for a large language model to respond in an appropriate manner. This process often requires iterative refinement and human insight to achieve optimal results. Based on current experiments conducted, trends in successful prompt engineering for our research include statements that would tell the large language model to reduce the complexity of responses like "as a beginner programmer". Other vital parts of the prompt include the code snippet, the warning generated by a static analysis tool, and a command. As the research evolves, the prompt may also change to handle different goals and methodologies to our study.

For our initial study, we have been using the prompt "Given this code and warnings produced by cppcheck/clang-check, can you (1) Explain why the code causing this warning should be fixed and (2) explain how to fix the warning.  Your answer should be formatted as "Explanation: ..., Fix: ..." and both the explanation and fix should be a single paragraph."  We provide a specific format for the large language model to respond with to make it easier to integrate the responses into tool support within a development environment.  We limit the length of the large language model's response to keep its response focused on important details about this specific example and avoid generalities about bug classes.

\subsection{Language Models}

Our initial experiments have used both ChatGPT through its web interface and additional models through the \texttt{ollama} tool \cite{ollama}.  \texttt{ollama} is run on a server with dual NVIDIA V100 GPUs.  For the purpose of our research, we are testing the capabilities of each model to determine which is most suited for our needs. So far, we have tested codegemma, gemma, llama3, and codellama.  Each one is trained up to billions of parameters and provides unique results.  We intend to continue evaluating these models on this problem as we progress with this research. To date, we have found the results from codellama to be the most promising due to its speed and expert understanding of code.  However, larger models, such as ChatGPT-4o and llama3-70b have also been highly successful at identifying the critical vulnerabilities in our test cases given the warnings produced by a static analysis tool.

\section{Evaluation}

We intend to provide the foundation for an upcoming user study on the effectiveness of using large language models to improve developer compliance with fixing potential bugs based on static analysis warnings.  This work moves toward determining if these hypotheses are valid, but further work will be required to determine if developers are actually motivated to fix potential bugs with the help of large language models.

\begin{itemize}
    \item \textbf{H1} Static analysis tools detect issues that lead to larger bugs in code.  Warnings are indicative of bigger problems.
    \item \textbf{H2} Large Language Models can provide context on static analysis warnings to persuade developers to fix warnings.
    \item \textbf{H3} Large Language Models can suggest reasonable solutions to quickly fix static analysis warnings.
\end{itemize}

In this paper, we make progress toward answering these questions, but we do not expect to provide complete answers until we have completed a user study grounded in these initial results.  We characterize the static analysis warnings generated by cppcheck and clang-check on a set of vulnerability test cases in the NIST Juliet suite.  We then delve into two case studies of applying large language models to static analysis warnings in a divide-by-zero vulnerability and a buffer overflow vulnerability.

\subsection{Characterization of Static Analysis Warnings}

\begin{table}
\caption{Warnings identified by Static Analysis Tools in NIST Juliet test cases.  Numbers shown are test cases in which at least one static analysis check generated a warning and the percentage of total test cases.}
\label{tab:warnings}
\begin{tabular}{c|cc}
\textbf{Bug CWE and Name} & \textbf{cppcheck} & \textbf{clang-check} \\ \hline
 CWE121 Stack Based Buffer Overflow & 3,876 (49\%) & 1,965 (25\%) \\ \hline
 CWE122 Heap Based Buffer Overflow & 2,743 (28\%) & 1,256 (13\%) \\ \hline
 CWE369 Divide by Zero & 846 (57\%) & 580 (39\%) \\ \hline
 CWE416 Use after Free & 274 (52\%) & 292 (56\%) \\ \hline
 CWE457 Use of Uninitialized Variable & 565 (52\%) & 568 (52\%) \\ \hline
\end{tabular}
\end{table}

We collected data on the number of NIST Juliet test cases that cause cppcheck and clang-check to generate warnings, shown in Table \ref{tab:warnings}.  In total, cppcheck generated warnings on 40\% of all test cases, and clang-check generated warnings on 23\% of all test cases.  These warnings do not directly identify the vulnerability indicated by the CWE description, but many of them identify issues in the code that are linked to the vulnerability.  For example, a data buffer that has been allocated to be of the correct size for a memory write may not be used by the code and will therefore be flagged as an unused variable.  In the next few subsections, we examine a few of these cases in more detail.

\subsection{Case Study: Divide-by-Zero}

\begin{figure}
    \centering
    \begin{lstlisting}[language=C++,frame=single]
#include <iostream>
using namespace std;

void f1() {
    int result = (100/0);
    cout << result << endl;
}

int main(int argc, char * argv[]) {
    f1();
    return 0;
}
\end{lstlisting}
\caption{Divide-by-zero test case - In this simple test case, the divide-by-zero always occurs, triggering a warning from the static analysis tool.}
    \label{fig:buffer}
\end{figure}

Our first case study involves the CWE369 Divide by zero. This is a simple and obvious mistake that triggers a static analysis tools such as cppcheck and clang-check to respond. Dividing by zero is mathematically undefined and as such, is also undefined in code.  

For this simple case, we used codellama, a LLM designed to both generate and discuss code.  We prompted it to explain the bug and suggest a fix. The explanation provided correctly analyzed why divide by zero is a bug and listed potential solutions. Due to the simplicity of this case, divide by zero establishes a baseline of how we will attempt to prompt LLM's in more complex cases.

codellama explains that "Division by zero is undefined in computer programming because it leads to inconsistent results, incorrect calculations, and even crashes the program.  For instance, when the denominator of a fraction is 0, the result of the division operation is not defined.  It can also cause a divide-by-zero error or floating point exception that stops the program's execution."  codellama suggests checking the input values to the division operator to ensure that the denominator is not zero to fix the bug.

In the more complex cases found in the Juliet test suite, a divide-by-zero error might manifest as a redundant assignment, an issue with variable scope, or as a condition known to be always true.  Further testing will be required to determine if large language models can still identify the critical vulnerabilities from related static analysis warnings.

\subsection{Case Study: Stack Based Buffer Overflow}

\begin{figure}
    \centering
    \begin{lstlisting}[language=C,frame=single]
#include <stdio.h>

int global = 5;

void f1() {
    char * data;
    char dataBuffer1[50];
    char dataBuffer2[100];
    if(global==5) {
        data = dataBuffer1;
    }
    data[0] = '\0';
    char source[100];
    memset(source, 'C', 100-1);
    source[100-1] = '\0';
    strcpy(data, source);
}

int main(int argc, char * argv[]) {
    f1();
    return 0;
}
\end{lstlisting}
\caption{Buffer Overflow Test Case - If global is set to 5, the smaller data buffer will be used by the strcpy function, leading to a buffer overflow.  In other cases, the variable data is uninitialized, also leading to a buffer overflow.}
    \label{fig:buffer}
\end{figure}

\begin{figure}
    \centering
    \includegraphics[width=.48\textwidth]{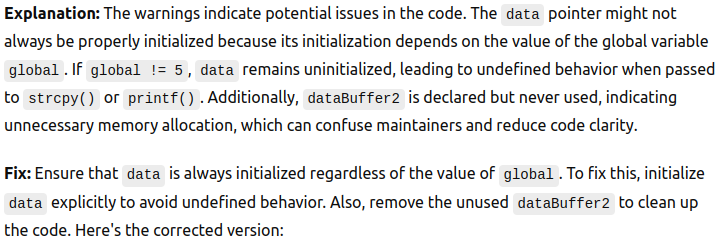}
    \caption{ChatGPT Output regarding the buffer overflow test case with both cppcheck warnings as additional input.}
    \label{fig:chatgpt_buffer}
\end{figure}

Our second case study involves a CWE121 Stack Based Buffer Overflow test case.  The code for this test case is provided in \ref{fig:buffer}.  As shown, a buffer overflow error can occur when \texttt{dataBuffer1} is used as the destination for the \texttt{strcpy} operation because it is smaller than the \texttt{source} array.  cppcheck detects that \texttt{dataBuffer1} and \texttt{dataBuffer2} are potentially unused, and it detects that \texttt{data} may be uninitialized.  Likewise, clang-check detects that the argument to \texttt{strcpy} may be uninitialized.

Figure \ref{fig:chatgpt_buffer} shows ChatGPT 4o's output when asked to explain and suggest a fix for the bug.  The explanation correctly describes the issue in the code and indicates why it is important to fix the code that causes these warnings.  Of particular interest, ChatGPT is able to incorporate an understanding of how static analysis must make assumptions about potentially unknown values of values in global variables, leading to this warning.  The fix suggested by ChatGPT successfully prevents the buffer overflow in this test case.

\section{User Study Design}

To evaluate the impact of large language model (LLM)-augmented warning messages on novice programmers, we have designed a controlled user study that compares participants’ responses to standard static analysis warnings versus those enhanced by a generative AI. The core objective of the study is to assess whether LLM-augmented warnings improve participants’ ability to identify and fix bugs and whether these enhancements influence their confidence and willingness to address flagged issues. Participants will be drawn from the U.S. Naval Academy’s undergraduate population, specifically targeting midshipmen with some programming experience.

The study will be structured as a between-subjects experiment. Participants will be randomly assigned to one of two groups: the control group will receive traditional static analysis warnings, while the experimental group will receive those same warnings augmented with plain-language explanations and suggested fixes generated by a large language model. Each participant will be tasked with debugging several short code examples, each containing one or more known issues detectable by static analysis tools. The examples will be drawn from educational and real-world sources to reflect a range of warning severities. Participants’ bug-fix success rates, time to completion, and confidence ratings will be recorded and analyzed.

In addition to quantitative measures of accuracy and efficiency, participants will complete a post-task survey capturing qualitative data, such as perceived usefulness of the warning messages and trust in the provided explanations. Demographic data will also be collected to explore any correlations between background and performance, including an analysis of whether LLM-augmented explanations have a disproportionately positive effect on underrepresented groups in computing. All data will be anonymized, and study procedures will be reviewed and approved by the Institutional Review Board. This study aims to inform the design of more effective warning and error reporting systems for educational and professional programming environments. We are currently waiting on approval by our Institutional Review Board to conduct this study.

\subsection{Automatic Test Case Extraction}

To support our user study with realistic, focused examples of buggy code, we developed a pipeline for automatically extracting compact test cases from the NIST Juliet Suite using a large language model (LLM). The Juliet Suite is a comprehensive collection of C/C++ programs designed to demonstrate a wide range of known software vulnerabilities. However, the Juliet examples include extensive scaffolding and multiple control paths, which can obscure the specific code responsible for triggering a static analysis warning. Furthermore, only a portion of the test cases exhibit a warning at all, and manually extracting a test case from the original code often lead to the warning not being present in the extracted code. Thus, we moved toward an automated approach. Our goal was to isolate these core buggy regions into standalone test cases that preserved the warning behavior while being short and readable enough for use in a controlled study.

The extraction process began by running each Juliet test case through clang-check and cppcheck, both widely-used static analysis tools. We captured the warning messages produced and then prompted an LLM to identify the smallest possible code snippet within the file that caused the warning. In addition to locating the buggy code, the LLM was asked to rewrite it into a standalone test case that preserved the original behavior. Variable names and function identifiers were obfuscated in the process to minimize bias in downstream studies and to prevent participants from inferring intent based on naming conventions.

To validate the fidelity of the generated test cases, each output was re-run through the same static analysis tool (either clang-check or cppcheck) to confirm that it still produced the original warning. This step ensured that the LLM had correctly isolated the relevant logic without removing or unintentionally correcting the underlying bug. At the time of writing, our pipeline had successfully produced 44 such validated test cases, with minimal human intervention, using the llama3 model with 70 billion parameters via ollama. This approach demonstrates how generative AI can be leveraged not only for augmenting warnings, but also for curating high-quality educational materials and evaluation artifacts from large, complex codebases.

\section{Discussion and Future Work}

The empirical results and findings from our case studies provide sufficient motivation to further study hypotheses H1, H2, and H3.  Large language models are able to analyze warnings related to the larger issue, explain why these warnings are relevant, and suggest a reasonable fix for the problem.

Once we have received approval from the Institutional Review Board, we intend to perform a user study to further test our hypotheses.  Using the test cases detailed in our case studies and additional test cases that have been automatically extracted using large language models, we will study and measure the behavior of undergraduate programmers confronted with static analysis warnings.  Details of the proposed user study are provided in the previous section.

To make this system usable, we intend to integrate enhanced warning messages into the Eclipse integrated development environment.  Eclipse provides support for integrating static analysis tools into a developer's workflow, and we will augment those tools to enable passing the output of the static analysis tool to a crafted prompt to a large language model, allowing the developer to work seamlessly with this system.

\subsection{Conclusion}

We have established a foundation for studying the impact of plain language explanations of static analysis warnings on developers' motivation to address these warnings, which may indicate serious underlying vulnerabilities. Initial results and case studies provide promising evidence supporting our hypotheses regarding the application of large language models in this context. To build on this foundation, we plan to conduct a comprehensive user study with undergraduate programmers to rigorously evaluate the validity of our hypotheses.

\section*{Acknowledgment}

We thank the anonymous reviewers for their insights that have been helpful in revising this paper.

\end{document}